# Ergodic capacity evaluation of wireless THz fiber extenders

Evangelos N. Papasotiriou, Alexandros-Apostolos A. Boulogeorgos, and Angeliki Alexiou
Department of Digital Systems, University of Piraeus, Piraeus, Greece
vangpapasot@unipi.gr , al.boulogeorgos@ieee.org, alexiou@unipi.gr

*Abstract*— **This paper focuses on delivering quantified results for the evaluation of the aggregated impact of stochastic antenna misalignment, multipath fading and hardware imperfections on the terahertz (THz) wireless fiber extenders. In this line, we present the appropriate signal model that accommodates the different technical and environmental parameters. In particular, it takes into consideration the antenna gains, the central frequency, the transmission range, the environmental conditions, i.e. temperature, humidity and pressure, the spatial jitter between the transmitter and receiver antennas, which results to antenna misalignment, the intensity of hardware imperfections, and the stochastic behavior of the wireless channel. Based on this model, we assess the joint impact of antenna misalignment and multipath fading, by providing Monte Carlo simulation results for the channels ergodic capacity.**

*Keywords—wirelles THz fiber extenders; hardware impairements; multipath fading, misalignment fading, ergodic capacity*

## I. INTRODUCTION

As the wireless systems evolve towards the fifth generation (5G), several technological advances, such as high order modulation schemes, massive multiple-input multiple-output (MIMO) and full duplexing, have been considered as possible enablers [1]. Although, these approaches provide significant performance improvement, there is still a lack of efficiency in handling the required quality of service and experience data driven oriented services [2], [3]. These observations have aspired the investigation of the THz band [4] – [7]. Communications in the THz band can provide an unprecented bandwidth increase and support extremely-high data rates at a cost of severe path attenuations, transceivers antenna misalignment and hardware imperfections [8]. The path attenuation originates from the high frequencies of this band, which cause interaction and energy absorption by the molecules of the propagation medium [9]. As a result, there exist plenty of published works regarding the modeling of the THz channel particularities [10], evaluating their impact on the THz system's performance [11], [12] and proposing solutions [13] – [18]. Furthermore, the use of high directive antennas on the transceivers causes antenna misalignment [19]. Finally, the hardware imperfections are the result of components mismatch and manufacturing defects in the radio frequency (RF) transceiver chain [20].

From the implementation point of view, in the THz band, the direct conversion architecture (DCA) is of much hype, due to its low-complexity and cost-effective configuration [21] – [25]. The main disadvantage of DCA is that it is sensitive to hardware imperfections, like in phase and quadrature imbalance (IQI), phase noise (PHN) and amplifier non linearities (ANL) [26], [27]. The effect of hardware imperfections was studied in several publications (see for example [28] – [41] and references therein), which concluded that they can significantly constrain the system's performance. However, their impact has been overlooked in the vast majority of THz systems technical literature. Only very recently, it was experimentally reported in [22], [25]. In more detail, in [25], the impact of hardware imperfections in the 300 GHz band was reported, while in [22], the authors highlighted their detrimental effect in THz wireless fiber extender systems.

Motivated by the above, in this paper, we present simulation results that quantify the aggregated effect of hardware imperfections in THz wireless extenders in the presence of stochastic antenna misalignment and small-scale fading. Specifically, after establishing the appropriate system model that considers the different technical and environmental parameters as well as the THz wireless channel characteristics, we assess the joint impact of misalignment and multipath fading in terms of ergodic capacity by delivering Monte Carlo simulation results.

## II. SYSTEM MODEL

We consider a THz wireless fiber extender equipped with highly directive antennas at both the TX and RX, to confront the severe channel attenuation. The employed system and channel model were initially presented in [52], where it is assumed that the complex information signal, $x$, is transmitted to the receiver, over a complex flat fading channel $h$ with complex additive noise $n$. The baseband equivalent received signal can be expressed as

$$y_i = hx + n \qquad (1)$$

where $h$, $x$ and $n$ are statistically independent. Additionally, $n$ is modeled as a complex zero mean additive white Gaussian process with variance $N_o$. Despite the fact that the received signal model presented in (1), accommodates the impact of the wireless channel and noise, the effect of hardware RF transceivers imperfections, namely IQI, PHN, as well as ANL, which, in high data rate systems, is detrimental [3], [26]. These imperfections generate a distortion between the intended signal $x$ and what is actually emitted and distort the received signal during the reception processing. To accommodate their influence at a given flat fading channel, we employ a generalized signal model [26], [42], which has been both theoretically and experimentally validated [43] – [46]. Based on this model, the baseband equivalent received signal can be written as

$$y = h(x + n_t) + n_r \quad (2)$$

In (2), $n_t$ and $n_r$ are respectively the distortion noises from the hardware imperfections at TX and RX [42], which can be modeled as [42], [47]

$$n_t \sim CN(0, k_t^2 P) \text{ and } n_r \sim CN(0, k_r^2 P|h|^2) \quad (3)$$

where $k_t$ and $k_r$ are non-negative parameters that determine the level of hardware imperfections at the TX and RX, respectively, while $P$ stands for the average transmitted power. The channel coefficient, $h$ can be obtained as

$$h = h_l h_p h_f \quad (4)$$

where $h_l$, $h_p$ and $h_f$ respectively model the path gain, the antenna misalignment fading and the fading $h_f$. The path gain coefficient can be expressed as

$$h_l = h_{fl} h_{al} \quad (5)$$

where $h_{fl}$ models the propagation gain and $h_{al}$ the molecular absorption gain. The term $h_{fl}$ is modeled by employing the Friis equation. Additionally, $h_{al}$ denotes the molecular absorption gain and can be evaluated as [48], [49]. The molecular absorption gain depends on the operational frequency, transmission distance and environmental conditions. The antenna misalignment, $|h_p|$ can be modeled as a stochastic process with probability density function (PDF) that can be obtained as [50]

$$f_{h_p}(x) = \frac{\gamma^2}{A_o^{\gamma^2}} x^{\gamma^2-1}, 0 \leq x \leq A_o \quad (6)$$

with $w_{eq}$ being the equivalent beam width radius at the RX. Moreover, $A_o$ is the fraction of the collected power when the TX and RX antennas are perfectly aligned. In order to accommodate the multipath fading effect, we model $|h_f|$ as a generalized $\alpha$ - $\mu$ distribution [51], with PDF that can be expressed as

$$f_{h_f}(x) = \frac{\alpha \mu^\mu}{\hat{h}_f^{\alpha\mu} \Gamma(\mu)} x^{\alpha\mu-1} \exp\left(-\mu \frac{x^\alpha}{\hat{h}_f^\alpha}\right) \quad (7)$$

where $\alpha > 0$, $\mu$ and $h_f$ stand for the fading parameter, normalized variance of the fading channel envelope and the $\alpha$-root mean value of the fading channel envelop, respectively.

## III. ERGODIC CAPACITY RESULTS

In this section, we investigate the joint effects of the deterministic and stochastic path-gain, i.e., misalignment and multipath fading, components as well as the impact of transceivers hardware imperfections in the ergodic capacity of the THz wireless fiber extender, which is defined as

$$C = E[\log_2(1+\rho)] \quad (8)$$

where $\rho$ represents the instantaneous signal-to-noise ratio (SNR) and $E[\cdot]$ returns the expected value, by illustrating respective simulation results. Unless otherwise stated, it is assumed that TX and RX gains are $G_t = G_r = 55$ dBi, $\alpha = 2$, $\mu = 1$ (corresponds to Rayleigh multipath fading, which is employed as a performance evaluation benchmark), $\mu = 4$ and $k_{tr} = k_t = k_r$ ( $k_{tr} = 0$ corresponds to the ideal RF-chain case, which is used here as a benchmark ). Moreover, standard environmental conditions, i.e., $\varphi=50$ %, $p = 101325$ Pa and $T = 296$ K are assumed.

Fig. 1 illustrates the ergodic capacity as a function of $\sigma_s$ for different levels hardware imperfections and values of $\mu$. The transmission distance is, $d = 30$ m, the operational frequency is set to $f = 275$ GHz and the transmitted signal power over the noise at the RX is $P/N_o = 25$ dB. As expected, for any given values of $\sigma_s$ and $k_{tr}$, the ergodic capacity for the curves having $\mu = 1$ is always lower than the respective ones with $\mu = 4$, because the latter represents multipath fading with a strong line-of-sight path component. Furthermore, we observe that for a given value of $\sigma_s$ and $\mu$, increasing $k_{tr}$, the ergodic capacity significantly decreases. For example, for $\sigma_s = 0.04$ m and $\mu = 4$ altering $k_{tr} = 0$ to $k_{tr} = 1$ yields ergodic capacity equal to 6.68 (bits/sec/Hz), 5.04 (bits/sec/Hz), 3.57 (bits/sec/Hz), 0.83 (bits/sec/Hz) and 0.58 (bits/sec/Hz), respectively. Additionally, for a given value of $\mu$ and $k_{tr}$, as $\sigma_s$ increases the ergodic capacity detrimentally decreases. As an example, for $\mu = 4$ and $k_{tr} = 0$, changing $\sigma_s = 0.01$ m to $\sigma_s = 0.1$ m the ergodic capacity degrades from 7.26 (bits/sec/Hz) to 4.15 (bits/sec/Hz).

In Fig. 2, the ergodic capacity is depicted as a function of $k_{tr}$ for different values of $\sigma_s$ and $\mu$. The transmission distance is set to $d = 20$ m, the operational frequency is $f = 300$ GHz and the transmitted signal power over the noise at the RX is $P/N_o = 20$ dB. As expected, for any given values of $\sigma_s$ and $k_{tr}$, the ergodic capacity for the curves having $\mu = 1$ is always lower than the respective ones with $\mu = 4$. Also, we observe that for any given value of $\sigma_s$ and $\mu$ as $k_{tr}$ increases, the ergodic capacity detrimentally decreases. For example, for $\sigma_s = 0.01$ m and $\mu = 4$ increasing $k_{tr} = 0$ to $k_{tr} = 0.2$ the ergodic capacity degrades from 7 (bits/sec/Hz) to 3.6 (bits/sec/Hz).

## IV. CONCLUSIONS

We evaluated the performance of wireless THz fiber extenders, under RF front-end hardware impairments, transceivers' antenna misalignment and multi-path fading. More specifically, we presented Monte Carlo simulation results for the assessment of the ergodic capacity. Our results reveal the detrimental effect of transceivers hardware imperfections and misalignment on the THz wireless system's performance, which are observed to be more severe compared to the effect of multipath fading. Furthermore, it can be observed that the RF imperfections are more significant than the misalignment, on the THz systems performance. Finally, the importance of accurate misalignment and impairments characterization for realistic performance assessment of THz wireless fiber extenders was highlighted.

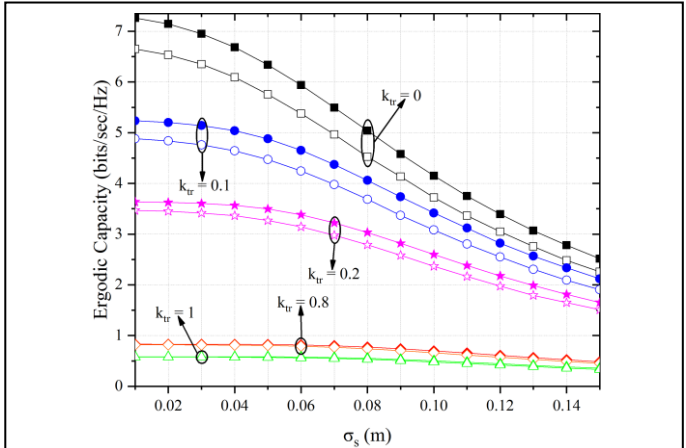

Fig. 1. Ergodic Capacity vs $\sigma_s$ for different levels of $k_{tr}$ and values of

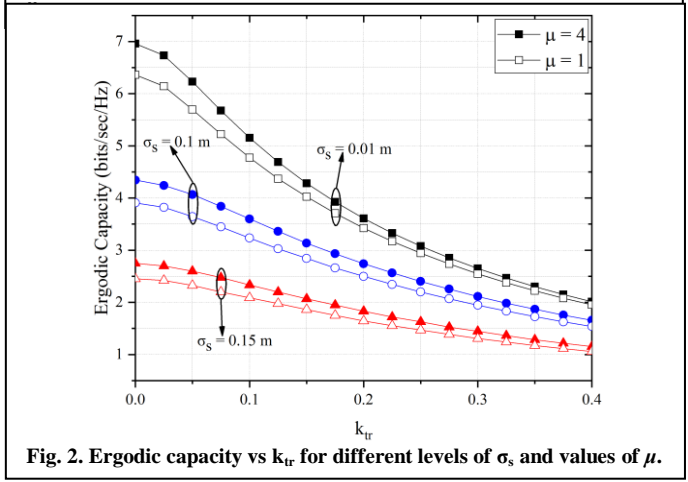

Fig. 2. Ergodic capacity vs $k_{tr}$ for different levels of $\sigma_s$ and values of $\mu$.

ACKNOWLEDGMENT

This work has received funding from the European Commission's Horizon 2020 research and innovation program TERRANOVA under grant agreement No. 761794.

REFERENCES

[1] M. Agiwal, A. Roy and N. Saxena, "Next Generation 5G Wireless Networks: A Comprehensive Survey," IEEE Communications Surveys Tutorials, vol. 18, pp. 1617-1655, 2016.

[2] A.-A. A. Boulogeorgos and G. K. Karagiannidis, "Low-cost Cognitive Radios against Spectrum Scarcity," IEEE Technical Committee on Cognitive Networks Newsletter, vol. 3, pp. 30-34, 11 2017.

[3] A.-A. A. Boulogeorgos, "Interference mitigation techniques in modern wireless communication systems," Thessaloniki, 2016.

[4] A.-A. A. Boulogeorgos, A. Alexiou, T. Merkle, C. Schubert, R. Elschner, A. Katsiotis, P. Stavrianos, D. Kritharidis, P. K. Chartsias, J. Kokkoniemi, M. Juntti, J. Lehtomäki, A. Teixeirá and F. Rodrigues, "Terahertz Technologies to Deliver Optical Network Quality of Experience in Wireless Systems Beyond 5G," IEEE Commun. Mag., vol. 56, pp. 144-151, 6 2018.R. Nicole

[5] H. Shokri-Ghadikolaei, C. Fischione, G. Fodor, P. Popovski and M. Zorzi, "Millimeter Wave Cellular Networks: A MAC Layer Perspective," IEEE Trans. Commun., vol. 63, pp. 3437-3458, 10 2015.

[6] I. F. Akyildiz, J. M. Jornet and C. Han, "Terahertz band: Next frontier for wireless communications," Phys. Commun., vol. 12, pp. 16-32, 9 2014.

[7] C. Zhang, "Breaking the blockage for big data transmission: Gigabit road communication in autonomous vehicles," IEEE Commun. Mag., vol. 56, no. 6, pp. 152-157, 2018.

[8] A.-A. A. Boulogeorgos, A. Alexiou, D. Kritharidis, A. Katsiotis, G. Ntouni, J. Kokkoniemi, J. Lethtomaki, M. Juntti, D. Yankova, A. Mokhtar, J.-C. Point, J. Machodo, R. Elschner, C. Schubert, T. Merkle, R. Ferreira, F. Rodrigues and J. Lima, "Wireless Terahertz System Architectures for Networks Beyond 5G," 2018.

[9] C. Han and Y. Chen, "Propagation Modeling for Wireless Communications in the Terahertz Band," IEEE Communications Magazine, vol. 56, pp. 96-101, 6 2018.

[10] G. A. Siles, J. M. Riera and P. G. Pino, "Atmospheric Attenuation in Wireless Communication Systems at Millimeter and THz Frequencies [Wireless Corner]," IEEE Antennas Propag. Mag., vol. 57, pp. 48-61, 2 2015.

[11] P. Boronin, V. Petrov, D. Moltchanov, Y. Koucheryavy and J. M. Jornet, "Capacity and throughput analysis of nanoscale machine communication through transparency windows in the terahertz band," Nano Commun. Networks, vol. 5, pp. 72-82, 9 2014.

[12] J. M. Jornet and I. F. Akyildiz, "Channel Capacity of Electromagnetic Nanonetworks in the Terahertz Band," in IEEE International Conference on Communications (ICC), Cape Town, South Africa, 2010.

[13] C. Lin and G. Y. L. Li, "Terahertz Communications: An Array-of-Subarrays Solution," IEEE Commun. Mag., vol. 54, pp. 124-131, 12 2016.

[14] K. Guan, G. Li, T. Kürner, A. F. Molisch, B. Peng, R. He, B. Hui, J. Kim and Z. Zhong, "On Millimeter Wave and THz Mobile Radio Channel for Smart Rail Mobility," IEEE Trans. Veh. Technol., vol. 66, pp. 5658-5674, 7 2017.

[15] A.-A. A. Boulogeorgos, E. N. Papasotiriou and A. Alexiou, "A distance and bandwidth dependent adaptive modulation scheme for THz communications," in 19th IEEE International Workshop on Signal Processing Advances in Wireless Communications (SPAWC), Kalamata, 2018.

[16] A.-A. A. Boulogeorgos, S. Goudos and A. Alexiou, "Users Association in Ultra Dense THz Networks," in IEEE International Workshop on Signal Processing Advances in Wireless Communications (SPAWC), Kalamata, 2018.

[17] N. Akkari, J. M. Jornet, P. Wang, E. Fadel, L. Elrefaei, M. G. A. Malik, S. Almasri and I. F. Akyildiz, "Joint physical and link layer error control analysis for nanonetworks in the Terahertz band," Wireless Networks, vol. 22, pp. 1221-1233, 01 5 2016.

[18] S. Han, I. C., Z. Xu and C. Rowell, "Large-scale antenna systems with hybrid analog and digital beamforming for millimeter wave 5G," IEEE Commun. Mag., vol. 53, pp. 186-194, 1 2015.

[19] A. S. Cacciapuoti, K. Sankhe, M. Caleffi and K. R. Chowdhury, "Beyond 5G: THz-Based Medium Access Protocol for Mobile Heterogeneous Networks," IEEE Commun. Mag., vol. 56, pp. 110-115, 6 2018.

[20] P. Rykaczewski, M. Valkama and M. Renfors, "On the Connection of I/Q Imbalance and Channel Equalization in Direct-Conversion Transceivers," #IEEE_J_VT#, vol. 57, pp. 1630-1636, 5 2008.

[21] A. J. Seeds, H. Shams, M. J. Fice and C. C. Renaud, "TeraHertz Photonics for Wireless Communications," J. Lightwave Technol., vol. 33, pp. 579-587, 2 2015.

[22] S. Koenig, D. Lopez-Diaz, J. Antes, F. Boes, R. Henneberger, A. Leuther, A. Tessmann, R. Schmogrow, D. Hillerkuss, R. Palmer, T.

Zwick, C. Koos, W. Freude, O. Ambacher, J. Leuthold and I. Kallfass, "Wireless sub-THz communication system with high data rate," Nat. Photonics, vol. 7, pp. 977 EP-, 10 2013.

[23] T. Nagatsuma, G. Ducournau and C. C. Renaud, "Advances in terahertz communications accelerated by photonics," Nat. Photonics, vol. 10, pp. 371 EP -, 5 2016.

[24] M. Elkhouly, Y. Mao, S. Glisic, C. Meliani, F. Ellinger and J. C. Scheytt, "A 240 GHz direct conversion IQ receiver in 0.13 μm SiGe BiCMOS technology," in IEEE Radio Frequency Integrated Circuits Symposium (RFIC), Seattle, 2013.

[25] I. Kallfass, I. Dan, S. Rey, P. Harati, J. Antes, A. Tessmann, S. Wagner, M. Kuri, R. Weber, H. Massler and others, "Towards MMIC-based 300GHz indoor wireless communication systems," IEICE Trans. Electron., vol. 98, pp. 1081-1090, 12 2015.

[26] T. Schenk, RF Imperfections in High-Rate Wireless Systems, The Netherlands: Springer, 2008.

[27] A. A. Abidi, "Direct-conversion radio transceivers for digital communications," #IEEE_J_JSSC#, vol. 30, pp. 1399-1410, 12 1995.

[28] A.-A. A. Boulogeorgos, V. M. Kapinas, R. Schober and G. K. Karagiannidis, "I/Q-Imbalance Self-Interference Coordination," IEEE Trans. Wireless Commun., vol. 15, pp. 4157-4170, 6 2016.

[29] A.-A. A. Boulogeorgos, P. C. Sofotasios, S. Muhaidat, M. Valkama and G. K. Karagiannidis, "The effects of RF impairments in Vehicle-to-Vehicle Communications," in IEEE 25th International Symposium on Personal, Indoor and Mobile Radio Communications - (PIMRC): Fundamentals and PHY (IEEE PIMRC 2015 - Fundamentals and PHY), Hong Kong, P.R. China, 2015.

[30] A.-A. A. Boulogeorgos, P. C. Sofotasios, B. Selim, S. Muhaidat, G. K. Karagiannidis and M. Valkama, "Effects of RF Impairments in Communications over Cascaded Fading Channels," #IEEE_J_VT#, vol. 65, pp. 8878-8894, 11 2016.

[31] A.-A. A. Boulogeorgos, H. B. Salameh and G. K. Karagiannidis, "On the Effects of I/Q Imbalance on Sensing Performance in Full-Duplex Cognitive Radios," in IEEE Wireless Communications and Networking Conference WS 8: IEEE WCNC'2016 International Workshop on Smart Spectrum (IWSS) (IEEEWCNC2016-IWSS), Doha, 2016.

[32] M. Mokhtar, A.-A. A. Boulogeorgos, G. K. Karagiannidis and N. Al-Dhahir, "OFDM Opportunistic Relaying Under Joint Transmit/Receive I/Q Imbalance," #IEEE_J_COM#, vol. 62, pp. 1458-1468, 5 2014.

[33] E. Björnson, M. Matthaiou and M. Debbah, "Massive MIMO systems with hardware-constrained base stations," in IEEE International Conference on Acoustics, Speech and Signal Processing (ICASSP), Florence, 2014.

[34] G. A., "Multi-channel energy detection under phase noise: analysis and mitigation," Mobile Networks and Applications, 2014.

[35] A. Gokceoglu, S. Dikmese, M. Valkama and M. Renfors, "Energy Detection under IQ Imbalance with Single- and Multi-Channel Direct-Conversion Receiver: Analysis and Mitigation," #IEEE_J_JSAC#, vol. 32, pp. 411-424, 3 2014.

[36] A.-A. A. Boulogeorgos, N. Chatzidiamantis, G. K. Karagiannidis and L. Georgiadis, "Energy Detection under RF impairments for Cognitive Radio," in Proc. IEEE International Conference on Communications - Workshop on Cooperative and Cognitive Networks (ICC - CoCoNet), London, 2015.

[37] L. Anttila, M. Valkama and M. Renfors, "Frequency-Selective I/Q Mismatch Calibration of Wideband Direct-Conversion Transmitters," IEEE Trans. Circuits Syst. II Express Briefs, vol. 55, pp. 359-363, 4 2008.

[38] L. Anttila, M. Valkama and M. Renfors, "Circularity-Based I/Q Imbalance Compensation in Wideband Direct-Conversion Receivers," #IEEE_J_VC#, vol. 57, pp. 2099-2113, 7 2008.

[39] E. Björnson, P. Zetterberg and M. Bengtsson, "Optimal coordinated beamforming in the multicell downlink with transceiver impairments," in IEEE Global Communications Conference, California, 2012.

[40] T. T. Duy, T. Q. Duong, D. Benevides da Costa, V. N. Q. Bao and M. Elkashlan, "Proactive Relay Selection With Joint Impact of Hardware Impairment and Co-Channel Interference," #IEEE_J_COM#, vol. 63, pp. 1594-1606, 5 2015.

[41] A. Gokceoglu, Y. Zou, M. Valkama, P. C. Sofotasios, P. Mathecken and D. Cabric, "Mutual Information Analysis of OFDM Radio Link Under Phase Noise, IQ Imbalance and Frequency-Selective Fading Channel," #IEEE_J_WCOM#, vol. 12, pp. 3048-3059, 6 2013.

[42] E. Björnson, M. Matthaiou and M. Debbah, "A New Look at Dual-Hop Relaying: Performance Limits with Hardware Impairments," IEEE Trans. Commun., vol. 61, pp. 4512-4525, 11 2013.

[43] C. Studer, M. Wenk and A. Burg, "MIMO transmission with residual transmit-RF impairments," in International ITG Workshop on Smart Antennas (WSA), 2010.

[44] M. Wenk, MIMO-OFDM Testbed: Challenges, Implementations, and Measurement Results, ETH, 2010.

[45] D. Dardari, V. Tralli and A. Vaccari, "A theoretical characterization of nonlinear distortion effects in OFDM systems," #IEEE_J_COM#, vol. 48, pp. 1755-1764, 10 2000.

[46] B. E. Priyanto, T. B. Sorensen, O. K. Jensen, T. Larsen, T. Kolding and P. Mogensen, "Assessing and Modeling the Effect of RF Impairments on UTRA LTE Uplink Performance," in IEEE 66th Vehicular Technology Conference, Baltimore, 2007.

[47] A.-A. A. Boulogeorgos, N. D. Chatzidiamantis and G. K. Karagiannidis, "Energy Detection Spectrum Sensing Under RF Imperfections," IEEE Trans. Commun., vol. 64, pp. 2754-2766, 7 2016.

[48] J. Kokkoniemi, J. Lehtomäki and M. Juntti, "Simplified molecular absorption loss model for 275-400 gigahertz frequency band," in 12th European Conference on Antennas and Propagation (EuCAP), London, 2018.

[49] A.-A. A. Boulogeorgos, E. N. Papasotiriou, J. Kokkoniemi, J. Lehtom"aki, A. Alexiou and M. Juntti, "Performance Evaluation of THz Wireless Systems Operating in 275-400 GHz Band," IEEE Vehicular Technology Conference (VTC), 2018.

[50] A. A. Farid and S. Hranilovic, "Outage Capacity Optimization for Free-Space Optical Links With Pointing Errors," Journal of Lightwave Technology, vol. 25, pp. 1702-1710, 7 2007.

[51] M. D. Yacoub, "The α-μ Distribution: A Physical Fading Model for the Stacy Distribution," IEEE Transactions on Vehicular Technology, vol. 56, pp. 27-34, 1 2007.

[52] A.-A. A. Boulogeorgos, E. N. Papasotiriou and A. Alexiou, "Analytical Performance Assessment of THz Wireless Systems," *IEEE Access*, vol. 7, pp. 11436-11453, 2019.